\documentclass{elsart}

\usepackage{graphicx}
\usepackage{amssymb}
\begin{document}
\begin{frontmatter}
\title
{Relativistic effects in the search for high density symmetry energy}
\author[TEX]{V.Greco\thanksref{gre}}, 
\author[LNS]{V. Baran\thanksref{bar}},
\author[LNS]{ M.Colonna},
\author[LNS]{M.Di Toro\corauthref{dit}}, 
\author[LNS]{T. Gaitanos\thanksref{theo}}
 and 
\author[MUN]{H.H. Wolter}

\corauth[dit]{ditoro@lns.infn.it}

\address[TEX]{Cyclotron Institute, Texas A\&M Univ., College Station, USA}%
\address[LNS]{Laboratori Nazionali del Sud INFN, Via S. Sofia 44,
I-95123 Catania, Italy\\
and Dipartimento di Fisica, Universit\`a degli Studi di Catania}%
\address[MUN]{Sektion Physik, 
 Universit\"at M\"unchen,\\ Am Coulombwall 1, D-85748 Garching, Germany}

\thanks[gre]{On leave from Laboratori Nazionali del Sud INFN, Catania, Italy
and Dipartimento di Fisica, Universit\`a degli Studi di Catania, partially
supported by a grant of the A.LaRiccia Foundation}
\thanks[bar]{On leave from NIPNE-HH and Bucharest University, Romania}
\thanks[theo]{On leave from Sektion Physik, 
 Universit\"at M\"unchen, Germany, supported by the German Ministry of 
Education and Science, BMBW, grant 06LM981.}

%\begin{document}

%\maketitle

\begin{abstract}
Intermediate energy heavy ion collisions open the unique
possibility to explore the Equation of State ($EOS$) of
nuclear matter far from saturation, in particular the
density dependence of the symmetry energy. Within a relativistic
transport model it is shown that the
isovector-scalar $\delta$-meson, which affects the high density
behavior of the symmetry energy density, influences the
dynamics of heavy ion collisions in terms of isospin
collective flows. The effect is largely enhanced by a relativistic
mechanism related to the covariant nature of the fields
contributing to the isovector channel. 
Results for reactions induced by $^{132}Sn$ radioactive beams
are presented.
The elliptic flows of nucleons and light isobars appear to be 
quite sensitive to microscopic structure of the symmetry term, 
in particular for particles
with large transverse momenta, since they represent an earlier
emission from a compressed source.
Thus future, more exclusive, experiments with relativistic radioactive
beams should be able to set stringent
constraints on the density dependence of the symmetry energy far
from ground state nuclear matter.
\end{abstract}


\end{frontmatter}

%\section{Introduction}
The high density behaviour of nuclear symmetry energy $E_{sym}(\rho_B)$
is very important for understanding many interesting astrophysical
phenomena, but it is absolutely not constrained by the predictions 
from several relativistic and non-relativistic
\cite{iso,enesim} models of nuclear matter.
The results can be roughly classified into two groups, i.e., one
where
the $E_{sym}(\rho_B)$ rises, {\it Asy-stiff}, and one in which it 
falls with increasing density, {\it Asy-soft} \cite{col98}. 
An increasing $E_{sym}(\rho_B)$ leads to a more proton-rich
neutron star whereas a decreasing one would make it more pure in
neutron content. As a consequence the chemical composition and 
cooling mechanism of
protoneutron stars \cite{latt91,sum94}, mass-radius
correlations \cite{pra88,eng94}, critical densities for kaon
condensation in dense stellar matter \cite{lee96,ku99}
as well as the possibility of a mixed quark-hadron phase \cite{kut00}
in neutrons stars will all be rather different.
It has recently been argued by means of simple thermodynamics
considerations
that the onset of the quark phase has a strong sensitivity  to the
behaviour of $E_{sym}(\rho_B)$ even for not very large asymmetries
\cite{dra02}.

The search for $E_{sym}(\rho_B)$ around saturation
density has driven a lot of theoretical and experimental efforts.
It seems to be rather
well established that heavy-ion collisions ($HIC$) at cyclotron energies
can give the possibility to extract some information on the symmetry
term
of the nuclear Equation of State ($EOS$) in region below and/or sligthly
above
the normal density \cite{col98,sca99,ko98,bao99,bar01,dit01}. 
On the other hand it is
quite desirable to get information on the symmetry energy at higher
density, where furthermore we cannot have complementary investigations
from nuclear structure like in the case of the
low density behaviour. Indeed $HIC$ provide a unique
way to create asymmetric matter at high density in terrestrial
laboratories.
Calculations within transport theory show that $HICs$ around $1AGeV$ allow to
reach a transient state of matter with more than twice the normal 
baryon density. Moreover, although the data are mostly of inclusive type
(and the colliding nuclei not very neutron rich), quite clearly a dependence
of some observables on charge asymmetry is emerging.

In this paper we show that a relativistic description of the nuclear
mean field can account for an enhancement of isospin effects during the
dynamics of heavy-ion collision. In particular future experiments with
radioactive
should be able to provide information on the vector part of isovector mean
fields
from collective flow analyses.

The isospin dependence of collective flows has been already discussed
in a non-relativistic framework \cite{sca99,bao00} using very different
$EOS$ with opposite behaviours of the symmetry term at high densities,
increasing repulsion ({\it asy-stiff}) vs. increasing attraction
 ({\it asy-soft}). The main new result shown here, in a fully 
relativistic scheme, is the importance at higher energies 
of the microscopic covariant structure 
of the effective interaction in the isovector channel: effective forces 
with very similar symmetry 
terms can give rise to very different flows in relativistic heavy ion
collisions.

We start from the Relativistic Mean Field ($RMF$) picture of the hadronic
phase of nuclear matter \cite{se86} which has been extensively used to
the study of the EOS for symmetric and asymmetric matter.
The $RMF$ describes an interacting system of nucleons (described as Dirac
spinors)
and meson classical fields.
The most common treatment of the isospin dependent part of the
interaction
is based on the introduction of an effective $\rho$-meson field
(vector-isovector) which can account for the known value of symmetry
energy at normal density.

However, a full description in a relativistic framework in principle
should
rely on the balance between a scalar and a vector field as stressed in
some
papers within the Hartree approximation \cite{ku97,liu02}, and as 
naturally
accounted for within the Dirac-Hartree-Fock ($DHF$)
\cite{prc_nuovo,gre01} or the Dirac-Brueckener-Hartree-Fock ($DBHF$)
\cite{hole01} schemes.
One could argue that the scalar-isovector $\delta$ meson $(a_0(980))$ 
is too heavy or
that the scalar field in the isoscalar channel may be due to the two pion
correlation, while
there is no equivalent possibility in the isovector channel.
On the other hand we stress that the vector-isovector field in the $RMF$
effective picture has not to be viewed as coming only from the exchange 
of the $\delta$ meson. In fact in the nuclear system contributions to the
isovector channel are mostly due to isoscalar mesons, as clearly
shown by DHF or DBHF scheme \cite{gre01,hole01,ma94,lop88}, through
important exchange and correlation terms.
Therefore when in the following we will refer to $\delta$ field 
indeed we mean a
$\delta-like$ field, i.e. the scalar isovector part of the effective
``interaction''.

In recent years some efforts have been devoted to the
effects of the scalar-isovector channel in finite nuclei, 
\cite{tywo99,fur00,bue02}.
Such investigations have not
shown a clear evidence for the $\delta$-field and this can
be understood considering that in finite nuclei one can test the
interaction properties mainly
below the normal density, where the effect of the $\delta-$channel on 
symmetry energy
and on the effective masses is indeed small \cite{liu02} and  eventually
could be absorbed into non linear
terms of the $\rho$ field. Moreover even studies of the asymmetric
nuclear matter by means of the Fermi Liquid Theory \cite{liu02} and a
linear response analysis have concluded that some properties, like the
borderline and the dynamical response inside
the spinodal instability region, are not affected by the
$\delta$ field \cite{gre03}.

Here we show that heavy-ion collisions around $1AGeV$ with
radioactive
beams can provide instead a unique opportunity to spot the presence of
the scalar isovector channel. In fact, due to the large counterstreaming
nuclear currents one may exploit the
different Lorentz nature of a scalar and a vector field.

%%%%%%%%%%%%%%%%%%%%%%%%%%
\begin{figure}
%\epsfysize=8.5cm
%\centerline{\epsfbox{esym-nl2rbuu.ps}}
\includegraphics[scale=0.75]{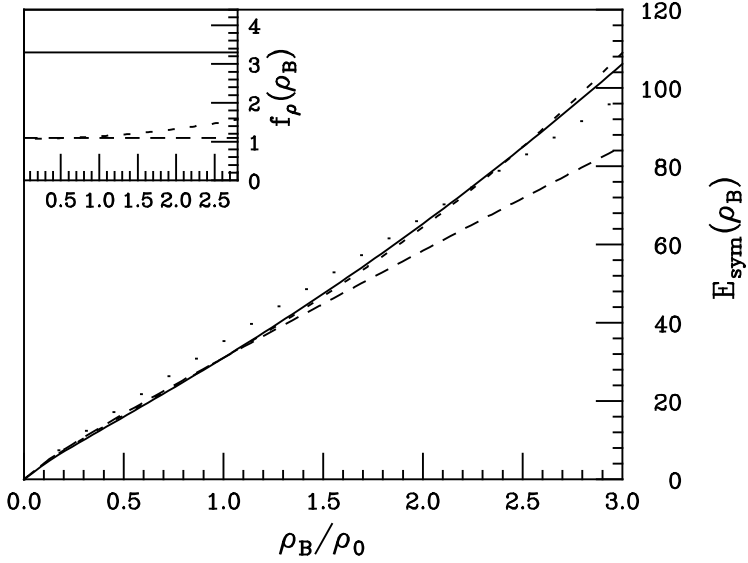}
\caption{Total (kinetic+potential) symmetry energy (in MeV) 
as a function of the
baryon
density.  Solid: $RMF-(\rho+\delta)$. Dashed: $RMF-\rho$. Short Dashed:
$RMF-D\rho$. In the insert
 the density behaviour of the $\rho$ coupling, $f_\rho$ (in $fm^2$), 
for the three
models is shown.}
\label{fig1}
\end{figure}
%%%%%%%%%%%%%%%%%%%%%%%%%%%

All models, including $RMF$, allow to use the parabolic approximation
for the description of the $EOS$ of asymmetric nuclear matter:
\begin{equation}\label{eq.1}
E(\rho_B,I)=E(\rho_B)+E_{sym}(\rho_B)\,I^2
\end{equation}
where $I=\frac{N-Z}{A}$ is the asymmetry parameter.
When both $\rho$-like and $\delta$-like channels are considered
$E_{sym}(\rho_B)$ can be written as \cite{liu02}:
\begin{equation}\label{eq.2}
E_{sym}(\rho_B) = \frac{1}{6} \frac{k_F^2}{E^\star_F} + \frac{1}{2}
\left[f_\rho - f_\delta
\left(\frac{M^\star}{E^\star_F}\right)^2\right] \rho_B\equiv
E_{sym}^{kin}+ E_{sym}^{pot}
\end{equation}
with $E^\star=\sqrt{k_F^2+M^{\star 2}}$, $M^\star$ the
effective Dirac mass and $f_{\rho,\delta}=(g_{\rho,\delta}/m_{\rho,\delta})^2$
are the coupling constants of the isovector channels.

We see that, when $\delta$ is included, the observed
$a_4=E_{sym}(\rho_0)$
value actually assigns  the combination
$[f_\rho - f_\delta (\frac{M^\star}{E^\star_F})^2]$
of the $(\rho,\delta)$ coupling constants, for futher details
see ref.\cite{liu02}.
In Fig.1 we report the density dependence of the symmetry energy
for three different models: one including only the
$\rho$ field ($RMF-\rho$), the other with ($\rho + \delta$)
fields ($RMF-(\rho+\delta)$) and the last, ($RMF-D\rho$), including only
$\rho$ field but with a covariant density dependence of $f_\rho$,
see ref. \cite{tywo99,flw95}. This is tuned just to give
at high density the same $E_{sym}(\rho_B)$ of the the ($RMF-(\rho+\delta)$)
case.
As shown in the following, the latter is useful for disentagling 
in the reaction dynamics the 
effects due to the
difference in $E_{sym}(\rho_B)$
from those directly linked to the strenght of the $\rho$ vector field.

Thus these models parametrize the isovector mean field either by only
the vector field with $f_{\rho}=1.1 fm^2$, or with a balance
between a vector field with $f_{\rho}=3.3 fm^2$ and a scalar one with
$f_{\delta}=2.4 fm^2$, or finally by
a normal density coupling $f_{\rho}(\rho_0)=1.1 fm^2$, but with
an increasing density dependence as shown in Fig.1 (insert).
We stress again that in $RMF-(\rho+\delta)$ the symmetry energy is coming
from a balance between a scalar attraction, ($\delta-like$),
and a vector repulsion, ($\rho-like$), which is now
roughly three times larger 
than in the $RMF-\rho$ case.

The choice of  $f_\delta$ is fixed relatively well
by DBHF \cite{hole01} and DHF \cite{gre01} calculations.
Therefore the effect described in the following is not artificially
enhanced, but based on a reliable estimate available at the
moment. In any case the aim of this work is to present some
qualitative new features expected in the reaction dynamics,
in particular for collective flows, just due to the introduction
of a scalar effective field in the isovector channel. 

Collective flows in heavy ion collisions give important information
on the dynamic response of excited nuclear matter
\cite{sto86,das93}. In particular the proton-neutron
differential flow $F^{pn}(y)$ \cite{bao00}
has been found to be a very useful probe of the isovector part of the $EOS$
sine it appears rather insensitive to the isoscalar potential
and the in medium nuclear cross section. 
The definition of the $F^{pn}(y)$ is
\begin{equation}\label{eq.3}
F^{pn}(y)\equiv \frac{1}{N(y)} \sum_{i=1}^{N(y)} p_{x_{i}} \tau_i
\end{equation}
where $N(y)$ is the total number of free nucleons at the
rapidity $y$, $p_{x_{i}}$ is the transverse momentum of
particle $i$ in the reaction plane, and $\tau_i$ is +1 and -1
for protons and neutrons, respectively.

For the theoretical description of heavy ion collisions we solve
the covariant transport equation of the Boltzmann type within the 
Relativistic Landau
Vlasov ($RLV$) method \cite{fuchs95} (for the Vlasov part) and applying
a Monte-Carlo procedure for the collision term. $RLV$ is a test particle
method using covariant Gaussians in phase space for the test particles.
The collision term includes elastic and inelastic processes involving
the production/absorption of the $\Delta(1232 MeV)$ and $N^{*}(1440
MeV)$ resonances as well as their decays into one- and two-pion channels.
Details about the used cross sections for all possible channels can
be found in Ref. \cite{hub94}.
An explicit isospin-dependent Pauli blocking term for the fermions is employed.
Asymmetry effects are suitably accounted for
in a self-consistent way with respect for the $RMF$ models
discussed above.

A typical result for the $^{132}Sn+^{132}Sn$ reaction at $1.5AGeV$ 
 (semicentral collisions) is
shown in Fig.2. The error bars are related to statistical fluctuations due
to the Monte-Carlo nature of the simulations.

We notice that the differential flow in case of the
$RMF-(\rho+\delta)$ (full circles and solid line) presents a stiffer
behaviour relative to the $RMF-\rho$ (open circles) model,
as expected from the more repulsive 
symmetry energy $E_{sym}(\rho_B)$ at high baryon densities, see Fig.1.
On the other hand it is quite
surprising that a relatively small
difference at $2\rho_0$ can result in a such different collective flows.
Indeed, we will see below that this is not the whole story.

We have repeated the calculation using the
$RMF-D\rho$ interaction, i.e. with only a $\rho$ contribution {\it but}
tuned to reproduce the same $EOS$ of the $RMF-(\rho+\delta)$ case.
The results, short-dashed curve of Fig.2, are very similar to the
ones of the $RMF-\rho$ interaction. 
Therefore we can explain the large flow effect as mainly due to 
the different strengths of the vector-isovector field 
between $RMF-(\rho+\delta)$
 and $RMF-\rho,D\rho$ in the relativistic dynamics. 
In fact if a source is moving the
vector field is enhanced (essentially by the local $\gamma$
Lorentz factor)\cite{nota}
relative to the scalar one. 

In order to get the idea we
write down, for an idealized situation, the ``force'' acting on a particle. 
From the isovector part of the interaction we get 
\begin{equation}
\frac{{d\vec p}^{\,*}_i}{d\tau}=\pm f_{\rho} \frac{E^*_i}{M^*_i}
\vec\nabla \rho_{3}
\mp f_\delta \vec\nabla \rho_{S3}
\label{eom}
\end{equation}
for protons and (upper signs), respectively neutrons (lower signs). 
${\vec p}^{\,*}$ is the
effective momentum, $\tau$ the particle proper time (averaged) and
$\rho_{3}=\rho_p-\rho_n$ the isovector baryon density (correspondingly
$\rho_{S3}$ the scalar one).
Here we have simplified the problem neglecting the contribution from 
the current gradient in the transverse direction and 
the current derivative with respect to time.
We are interested in the difference between the force acting on 
a neutron and on a proton, respectively.
Oversimplifying the $HIC$ dynamics we consider locally neutrons and  protons
with the same $\gamma$ factor (i.e. with
the same speed). Then Eq.\ref{eom} can be expressed approximately
by the following
transparent form ($\rho_{S3}=\frac{M^*}{E^*}\rho_{3}$):
\begin{equation}
\frac{{d\vec p}^{\,*}_p}{d\tau}-\frac{{d\vec p}^{\,*}_n}{d\tau}\simeq 2\left
[\gamma f_{\rho}
- \frac{f_{\delta}}{\gamma}\right]\vec\nabla \rho_{3}
\end{equation}
where $\gamma$ is the Lorentz factor for the collective motion 
of a given ideal cell.

Keeping in mind that
$RMF-(\rho+\delta)$
has a three times larger $\rho$ field it is clear that dynamically
the vector-isovector
mean field acting during the $HIC$ is much greater than the one of the
$RMF-\rho,D\rho$ cases.
Then the isospin effect is mostly caused by 
the different
Lorentz structure of the
``interaction'' which results in a dynamical breaking of the balance
between
the $\rho$ vector and $\delta$ scalar fields, present in nuclear matter
at equilibrium.  This effect is analogous to the interplay between
the isoscalar vector- and scalar-fields which is seen in the
magnitude and energy dependence of the real part of the optical
potential, ref. \cite{gait01}.
%%%%%%%%%%%%%%%%%%%%%
\begin{figure}[ht]
%\epsfysize=6.5cm
%\centerline{\epsfbox{sn132_15b6-40.ps}}
\includegraphics[scale=0.75]{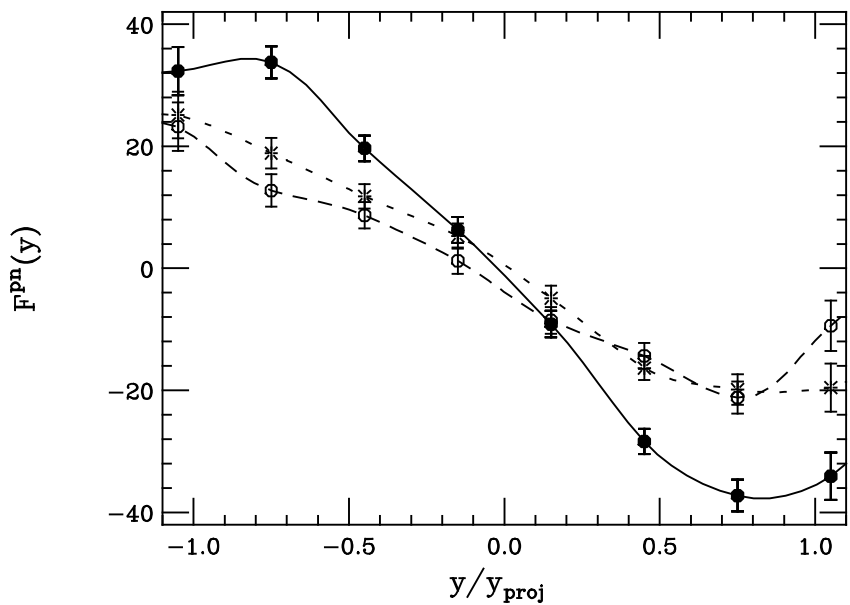}
\caption{ $^{132}Sn+^{132}Sn$
reaction at 1.5 AGeV (b=6fm): Proton-neutron transverse differential 
collective flow (in $MeV/c$), vs. rapidity, from the three different models 
for the isovector mean fields.
Full circles and solid line: $RMF-(\rho+\delta)$.
Open circles and dashed line: $RMF-\rho$.
Stars and short dashed line : $RMF-D\rho$.
}
\label{fig2}
\end{figure}
%%%%%%%%%%%%%%%%%%%%

In order to characterize the effect on differential collective flows we have
calculated
the slope $dF^{pn}(y)/d(y/y_{proj})$ at mid-rapidity.
Its value is $46.7MeV/c$ for $RMF-(\rho+\delta)$
and $23.4MeV/c$ for $RMF-\rho$, i.e. a factor two difference.

We have also performed some calculations at lower beam energies. We have
found that up to $500AMeV$ there is no valuable difference in the differential
flow predictions among the models
discussed here. The effect coming from the strength of
$\rho$ field  starts to become important around $1AGeV$, as expected from 
the relativistic mechanism.

%%%%%%%%%%%%%%%%%%%%%%%%
\begin{figure}[htb]
%\epsfysize=6.5cm
%\centerline{\epsfbox{sn132_15v2pnb6.ps}}
\includegraphics[scale=0.75]{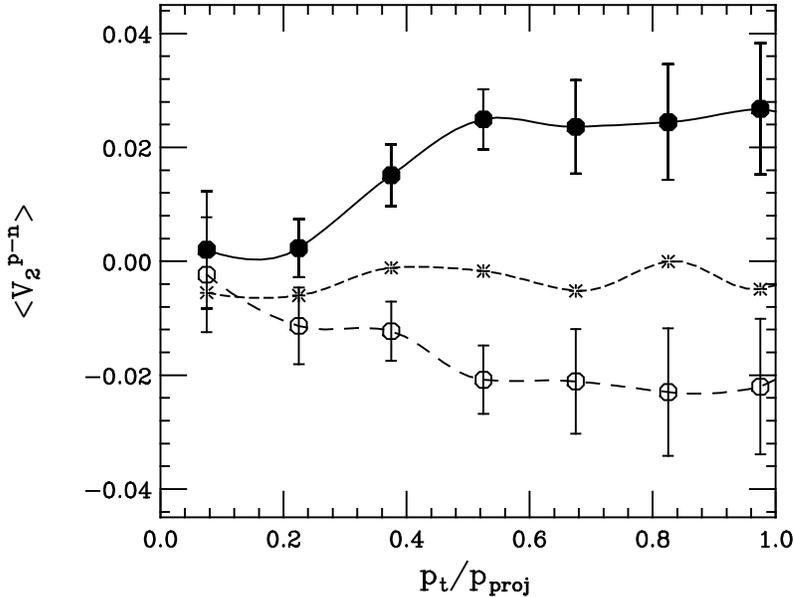}
\caption{Difference between neutron and proton elliptic flow
as a function of the transverse momentum in the
$^{132}Sn+^{132}Sn$ reaction at 1.5 AGeV b=6fm in the
rapidity range $-0.3 \leq y/y_{proj} \leq 0.3$.
Full circles and solid line: $RMF-(\rho+\delta)$.
Open circles and dashed line: $RMF-\rho$.
Stars and short dashed line: $RMF-D\rho$.
The error bars of the $RMF-D\rho$ curve are similar to the other cases and
are not shown just for comprehensibility.
}
\label{fig3}
\end{figure}
%%%%%%%%%%%%%%%%%%%%%%%%

Another interesting observable is the elliptic flow $v_{2}(y,p_t)$, which
is derived as the second coefficient from a Fourier expansion of the
azimuthal distribution
$N(\phi,y,p_{t})=v_{0}(1+v_{1}cos(\phi)+2v_{2}cos(2\phi))$. It can
be expressed as
 $$v_2=<\frac{p^2_x-p^2_y}{p^2_t}>$$
where $p_t=\sqrt{p^2_x+p^2_y}$ is the transverse momentum \cite{oil92,daniel}.

A negative value of $v_2$ corresponds to the emission of
matter
perpendicular to the reaction plane, $squeeze-out$ flow. 
The $p_t$-dependence of
$v_2$,
 which has been recently investigated by various groups
\cite{daniel,eflow,gait01}, is
very sensitive to the high density behavior of the $EOS$ since highly
energetic
particles ($p_t \ge 0.5$) originate from the initial compressed and
out-of-equilibrium phase of the collision, see e.g. Ref. \cite{gait01}.

In Fig.3 we present the $p_t$ dependence of
the proton-neutron difference of the elliptic flow in the same
very exotic  $^{132}Sn+^{132}Sn$ reaction at $1.5AGeV$ 
 (semicentral collisions), for mid-rapidity emissions. 
The larger error bars correspond to a reduced
statistics when a selection on different $p_t$ bins is introduced.
The effect is increasing for larger $p_t$ values due to the 
smaller number of contributions.

From Fig.3 we see that, in spite of the statistical errors, 
in the $(\rho+\delta)$
dynamics the high-$p_t$ neutrons show a much larger $squeeze-out$.
This is fully consistent with an early emission (more spectator shadowing)
due to the larger repulsive $\rho$-field. We can expect this appreciable 
effect since the relativistic enhancement discussed above is
relevant just at the first stage of the collision.
The $v_2$ observable, which is a good {\it chronometer} of the reaction
dynamics, appears to be particularly sensitive to the Lorentz structure
of the effective interaction.

We have repeated the same set of simulations for the more realistic
$^{132}Sn+^{124}Sn$
reaction at 1.5 AGeV (b=6fm), that likely could be studied with 
the new planned radioactive beam facilities at intermediate energies.
The results are shown in Fig.4. The effect of the different structure of the 
isovector channel is still quite clear, of course with a reduction
due to the smaller isospin density in the interaction region.
Particularly evident is again the splitting in the high $p_t$
region of the elliptic flow.
%%%%%%%%%%%%%%%%%%%%%%%%
\begin{figure}[htb]
\includegraphics[scale=0.75]{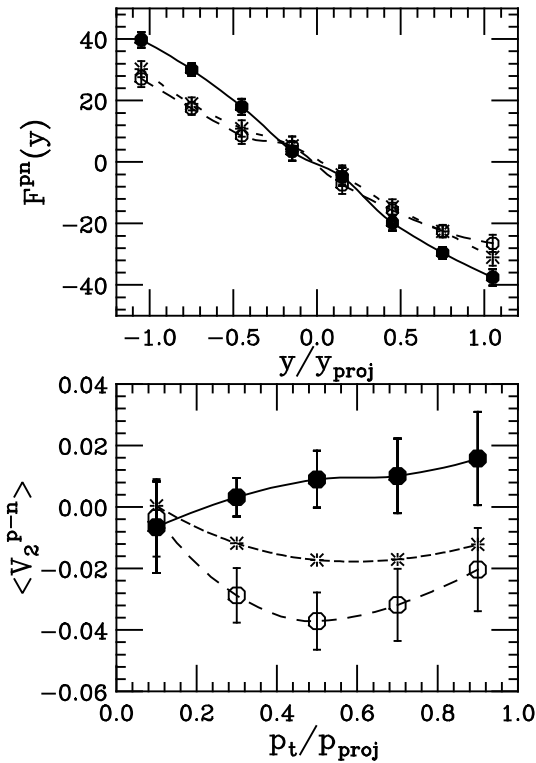}
\caption{$^{132}Sn+^{124}Sn$
reaction at 1.5 AGeV (b=6fm) from the three different models for the
isovector mean fields.
Top: as in Fig.2. Bottom: as in Fig.3.
Full circles and solid line: $RMF-(\rho+\delta)$.
Open circles and dashed line: $RMF-\rho$.
Stars and short dashed line: $RMF-D\rho$.
Error bars: see the text and the previous caption.
}
\label{fig4}
\end{figure}
%%%%%%%%%%%%%%%%%%%%%%%%

In conclusion intermediate energy heavy-ion collisions
with radioactive beams can give information on the symmetry energy 
at high baryon density and on its detailed microscopic structure. 
We have shown that such experiments provide
a unique tool to investigate the strength of the $\delta-like$ field.
The sensitivity is enhanced relative to the static property $E_{sym}(\rho_B)$ 
because of the more fundamental covariant nature of the fields
involved in $HIC$ dynamics.

Collective flows observables are found to be sensitive
to isospin effects. Due to the time selectivity on the emitted
particles the elliptic flow measurements appear to be the
most appealing, especially for nucleons and light isobars
at high transverse momentum.  

%%%%%%%%%%%%%%%%%%%%%%%%%%%%%%%%%%%%%

\end{document}